\documentclass{article}

\usepackage{arxiv}

\usepackage[T1]{fontenc}
\usepackage{lmodern}

\usepackage{graphicx}

\usepackage{cite}
\usepackage{amsmath,amssymb,amsfonts}
\usepackage{mathrsfs}
\usepackage{algorithmic}
\usepackage{textcomp}
\usepackage{xcolor}
\usepackage{optidef} 
\usepackage{bm}
\usepackage{subfigure}
\usepackage{stackengine}
\usepackage{tabularray}

\usepackage{hyperref}       

\begin{document}

\title{On Searching for Minimal Integer Representation of Undirected Graphs}

%

\newif\ifuniqueAffiliation
\uniqueAffiliationtrue

\ifuniqueAffiliation 


\author{ \href{https://orcid.org/0000-0001-7329-1468}{\includegraphics[scale=0.1]{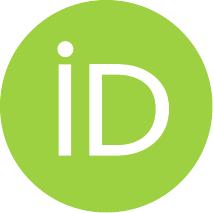}\hspace{0.5mm} Victor Parque}\thanks{Corresponding author.} \\
	Department of Modern Mechanical Engineering\\
	Waseda University\\
	3-4-1 Okubo, Shinjuku, Tokyo, Japan 169-8555 \\
	\texttt{parque@aoni.waseda.jp} \\
	\And
	\href{https://orcid.org/0000-0002-3371-6182}{\includegraphics[scale=0.1]{orcid.pdf}\hspace{0.5mm}Tomoyuki Miyashita} \\
	Department of Modern Mechanical Engineering\\
	Waseda University\\
	3-4-1 Okubo, Shinjuku, Tokyo, Japan 169-8555 \\
}



\maketitle              
\begin{abstract}
Minimal and efficient graph representations are key to store, communicate, and sample the search space of graphs and networks while meeting user-defined criteria. In this paper, we investigate the feasibility of gradient-free optimization heuristics based on Differential Evolution to search for minimal integer representations of undirected graphs. The class of Differential Evolution algorithms are population-based gradient-free optimization heuristics having found a relevant attention in the nonconvex and nonlinear optimization communities. Our computational experiments using eight classes of Differential Evolution schemes and graph instances with varying degrees of sparsity have shown the merit of attaining minimal numbers for graph encoding/representation rendered by exploration-oriented strategies within few function evaluations. Our results have the potential to elucidate new number-based encoding and sample-based algorithms for graph representation, network design and optimization.

\end{abstract}

\section{Introduction}

Graphs allow to compute with dependencies in several fields. Devising succinct graph representations is essential to realize efficient mechanisms for storing, transmission and sampling graphical structures ubiquitously. Although matrices and lists are useful data structures allowing the simple and straightforward representation of graphs, they are unable to meet the information-theoretic tight bounds. In a seminal paper, Turan outlined the concept of succinctness when dealing with graph representations, and proposed the succinct representations of both labeled and unlabeled planar graphs using size up to $12n$ bits ($n$ is the number of nodes in the graph)\cite{turan84}. And Farzan and Munro proposed the representation of arbitrary graphs with the size being a multiplicative factor $1+\epsilon$ above the information-theoretic minimum, for small positive $\epsilon$\cite{farza08}. Finding efficient representation of arbitrary graphs implies finding regularities in graphs that can be polynomially encoded/decoded. As such, the community has used regular features in the graph topology such as planarity\cite{turan84,jacobson89,barbay2012succinct,ale06,he99}, triangularity\cite{barbay2012succinct,ale06,he2000fast}, separability\cite{he2000fast,blelloch2010succinct,blandfor03} and symmetry \cite{katebi12}. When using regularity in graph representation, it is often useful find label encodings that allow tests for node adjacency\cite{erdos1989representations,gallian2018dynamic}. Generally speaking, when regularity in the graph topology is known, the search space of tailored graph representations is reduced; thus it becomes feasible to investigate tight bounds on the representation mechanism.

The community has also explored the feasibility of compressing real-world networks\cite{claude2010fast,boldi2011layered,buehrer2008scalable,brisaboa2014compact}, in which compression efficiency is often reported in bits per edge. It is also possible to realize the number-based representation of graphs by using combinatorial ideas\cite{kreher1999combinatorial,smc14,bigcomp17,compsac18}, in which a graph coding scheme computes an integer number representing the graph, and a decoding scheme generates the graph topology from the integer number. Representing graphs with numbers is advantageous when a-priori knowledge of graph regularity is unknown. Also, the number representing the graph meets the information-theoretic tight bound over all possible graph representations in the same class. On the other hand, when the problem is to find an optimal graph topology, the number based representation enables the one-dimensional (integer) search space and the implicit parallelization of search.

One of the key questions in number-based graph representation is whether exploring the landscape of enumerative encodings would render representations with fewer digits, and thus smallest possible integers. Although approaches for succinct graph encodings have relied on apriori knowledge of graph invariants such as planarity, separability, symmetry, and sparsity, the study of representing arbitrary graphs with the smallest possible integers has received little attention in the literature. In this paper, we study the feasibility of using gradient-free population-based optimization heuristics inspired by Differential Evolution for smallest/minimal integer representation of graphs, and present an analysis of the convergence performance. In particular, our contributions are as follows: (1) We formulate the problem of searching for minimal integer representation of graphs, portray examples of the search space, and evaluate the performance of eight gradient-free optimization algorithms with varying forms of exploration-exploitation based on Differential Evolution schemes. (2) Our computational experiments using graph instances with varying degrees of sparsity has shown the merit of exploration strategies to attain better convergence with few function evaluations.

The rest of this paper is organized as follows. Section 2 provides the preliminaries on enumerative and minimal graph representation, section 3 describes the computational experiments, and section 4 concludes our paper.

\section{Graphs with Minimal Integers}


\subsection{Enumerative Representation of Graphs}

Let $\mathscr{G}_{n,m}$ be the class of undirected graphs with $n$ nodes and $m$ edges. When graphs in the class $\mathscr{G}_{n,m}$ are represented by integer numbers $g$, the following condition holds: $g \in  [0, |\mathscr{G}_{n,m}| - 1]$, in which,

\begin{equation}\label{num}
  |\mathscr{G}_{n,m}|  = \binom{\binom{n}{2}}{m}.
\end{equation}

The above implies that graphs in the class $\mathscr{G}_{n,m}$ are combinatorial structures that involves the combinations of $\binom{n}{2}$ edges taken $m$ at the time. Let the graph instance $G = (V , E) \in  \mathscr{G}_{n,m}$, with $n = |V|$ and $m = |E|$; a numerical representation (or numerical encoding) of $G$ is the tuple $(g, ~CODE, ~DECODE)$ as shown by Fig. \ref{number}:

\begin{equation}\label{encode}
  CODE: G \rightarrow g,
\end{equation}

\begin{equation}\label{encode}
  DECODE(CODE(g)) \rightarrow G.
\end{equation}

\begin{figure}[t]
\centering
\includegraphics[width=0.65\textwidth]{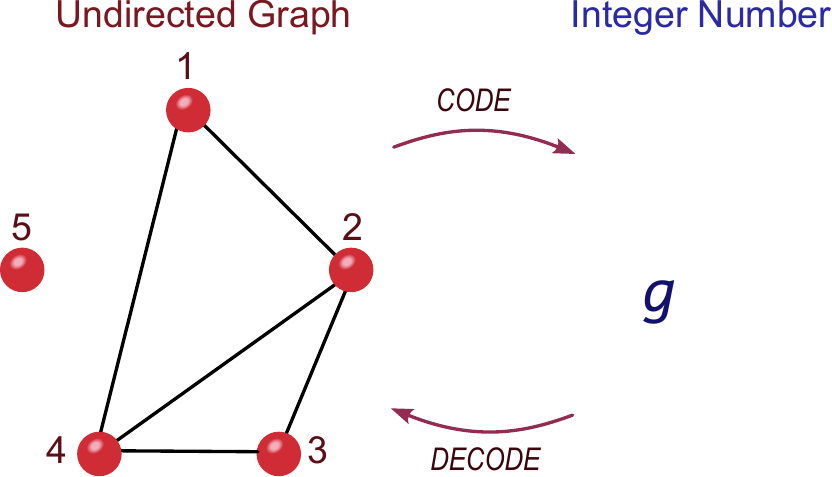}
\caption{Basic concept of representing graphs as numbers. Given an graph, the \emph{CODE} function generates the number $g$ representing the graph and, viceversa, the \emph{DECODE} function renders the graph from the integer number $g$.}
\label{number}
\end{figure}

For a defined labeling of nodes with the tuple $(1, ~2, ~3, ..., ~n)$, it is possible to define the \emph{CODE} function as the following algebraic relation\cite{kreher1999combinatorial,smc14}:

\begin{equation}\label{comnum}
  g(x) = \sum_{c = 1}^{m} (-1)^{m-c} \Bigg [ \binom{i_c}{c} - 1 \Bigg],
\end{equation}
where $i_c$ is the numerical label (integer number) of the $c$-th edge $(u, v) \in E$, computed as follows\cite{smc14}:

\begin{equation}
  i_c = \binom{u-1}{2} + v, ~ \text{for } u > v,
\end{equation}
where $u \in [1,n]$ and $v\in [1,n]$ are labels of nodes from the set $V$.

The reader may note that the number representation of a graph is contingent upon the values of $n$ and $m$, and the above-mentioned definitions consider a defined labeling order of the set of nodes $V$. It is possible to define the \emph{DECODE} function with a decoding algorithm based on the revolving door order\cite{kreher1999combinatorial,smc14}.


\subsection{Minimal Integer Representation}

Graphs can be represented with minimal integers by solving the following:

\begin{equation}\label{prob}
\begin{aligned}
& \underset{x}{\text{Minimize}}
& & L(x) \\
& \text{subject to}
& & x \in \mathbf{F}(n)
\end{aligned}
\end{equation}
where the objective function is

\begin{equation}\label{obj}
  L(x) = \log (g(x)),
\end{equation}
where $x$ denotes the representation of the ordering of labels in the graph, and $\mathbf{F}(n)$ is the search space related to the factorial numbering system of $n$. The problem described above is NP-hard, and without further knowledge of the graph topology or regularities in the structure, computing the gradient of $L$ is unfeasible; thus, the use of a gradient-free and stochastic optimization heuristic is a desirable choice to sample the search space of $F(x)$. The logarithmic function in Eq. (\ref{obj}) facilitates computing the number of digits in number-based representations of graphs, thus it provides an intuitive idea on the space needed to encode undirected graph. Furthermore, the use of the factorial (or \emph{factoradic}) space $\mathbf{F}(n)$ in Eq. (\ref{prob}) is due to the permutation of node labels in terms of $n!$. 

To exemplify the landscape of the above-mentioned objective function, we consider the following examples of graphs:

\begin{figure}[t]
\centering
\hfill
\subfigure[$G_1$, $n = 5, m = 5$]{\includegraphics[width=0.3\textwidth]{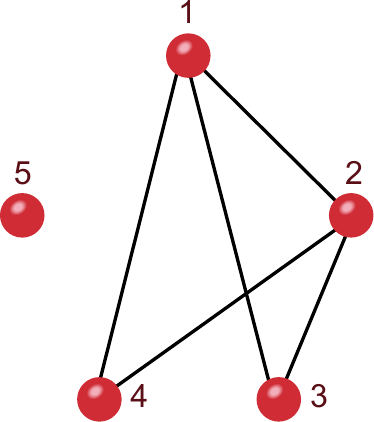}}
\hfill
\subfigure[$G_2$, $n = 5, m = 8$]{\includegraphics[width=0.3\textwidth]{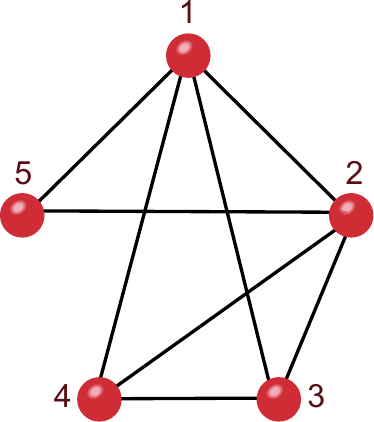}}
\hfill
\caption{Example of undirected graph topologies.}
\label{example}
\end{figure}

\begin{itemize}
  \item $G_1$: an undirected graph with $n = 5$ nodes and $m = 5$ edges, and
  \item $G_2$: an undirected graph with $n = 5$ nodes and $m = 8$ edges.
\end{itemize}

Fig. 2 shows the topologies of graphs $G_1$ and $G_2$. For both graphs, $n = 5$, yet the reader may note that a node with label 5 in $G_1$ is not connected to any other node. Then, one can compute the objective function $L$ over the factorial space $\mathbf{F}(n)$, for each permutation of the node labels. For $n = 5$, there exists $5! = 120$ possible orders of the labels, thus 120 possible values of the metric $L$.

\begin{figure*}[t]
    \centering
	\subfigure[Landscape of $L(x)$ for $G_1$, $n = 5, m = 5$]{\includegraphics[width=0.98\textwidth]{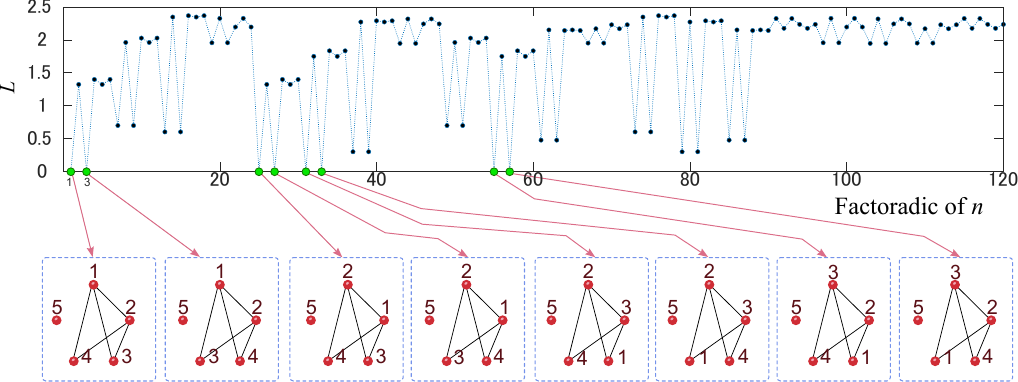}}
	\hfill
	\subfigure[Landscape of $L(x)$ for $G_2$, $n = 5, m = 8$]{\includegraphics[width=0.98\textwidth]{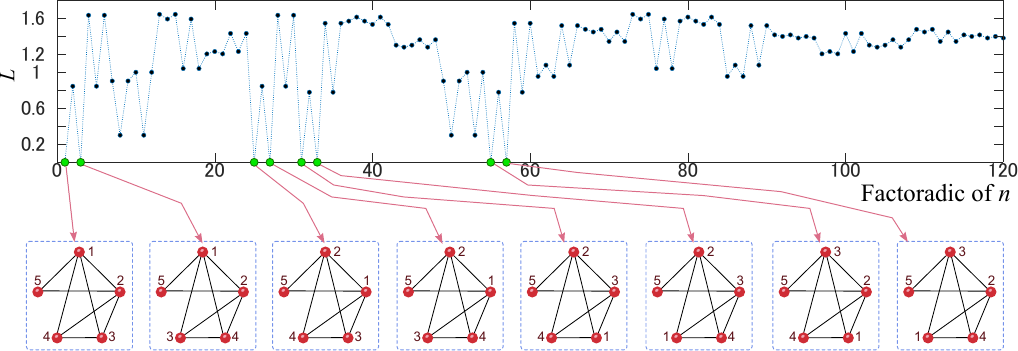}}
	\hfill
	\caption{Example of the landscapes of the fitness functions for representing of graphs by smallest numbers. Green marks imply the lowest smallest integers.}
	\label{example}
\vspace{-0.5cm}
\end{figure*}

%
%
%

In order to show the landscape of the objective function, Fig. 3 shows the rendering of the metric $L$ for both graphs $G_1$ and $G_2$ when evaluating all possible 120 permutations of labels. Here, the x-axis shows the factorial number system corresponding to the ordering of the label (for $n = 5$, the x-axis shows numbers in the range 1-120), and the y-axis shows the metric $L$. By observing Fig. 3, one can note that the landscape is nonsmooth, and it is possible to find regions with smallest values of $L$. Such regions (labeled with green color) are of interest to find node and edge labels that render smallest/minimal integer representation. Examples of such node labels are provided in Fig. 3-(a) and Fig. 3-(b), for each graph $G_1$ and $G_2$, respectively. Sampling through the factorial number system allows to explore possible configurations of graphs whose representation may lead to the use of fewer number of digits.

\begin{figure*}[t!]
    \centering
	\subfigure[Landscape of $L(x)$ for arbitrary graphs with $n = 5, ~m = 2, ...,8$]{\includegraphics[width=0.985\textwidth]{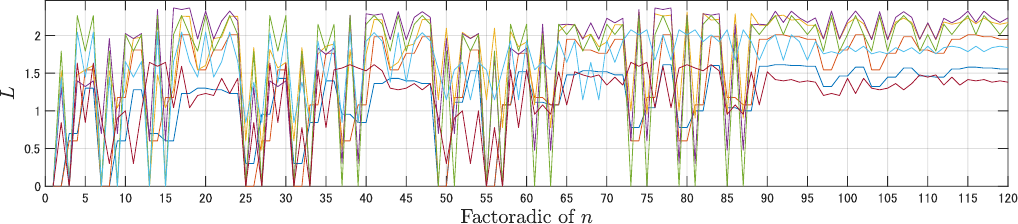}}
	\hfill
	\subfigure[Landscape of $L(x)$ for arbitrary graphs with $n = 6, ~m = 2,...,14$]{\includegraphics[width=0.985\textwidth]{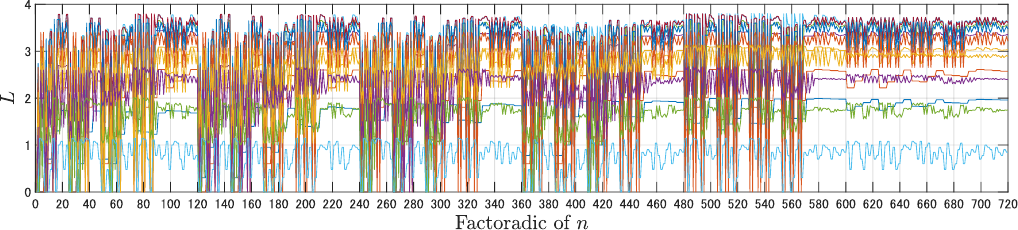}}
	\hfill
	\caption{Examples of landscapes of the function $L$ for arbitrary graphs. (a) The case of $n = 5, m \in [2, 8]$. (b) The case of $n = 6$ and $m \in [2,14]$.}
	\label{examplefull}
\vspace{-0.5cm}
\end{figure*}

In order to show the overall landscape for distinct and arbitrary number of edge configurations, Fig. \ref{examplefull} shows the landscapes of the metric $L$ for arbitrary graphs when considering varying number of edges. Here, we show the the case of undirected graphs with $n = 5, ~m \in [2, 8]$, and the case of undirected graphs with $n = 6$ and number of edges $m \in [2,14]$. Compared to Fig. 3, the landscape of in Fig. 4 shows the larger number of instances in which it is possible to attain smallest integer representation. This is due to the graph configurations arising from larger number of edges and nodes. Also, sparse undirected graphs with fewer edges compared to number of nodes will allow the larger number of combinatorial permutations with smallest integer representation. Furthermore, although the minimization of the objective function in Eq. (\ref{obj}) implies searching for smallest numbers, it is possible to explore other objective functions to search for graphs with tailored properties (e.g., by using factorization and the fundamental theorem of arithmetic).

\begin{figure*}[t]
\centering
\includegraphics[width=1\textwidth]{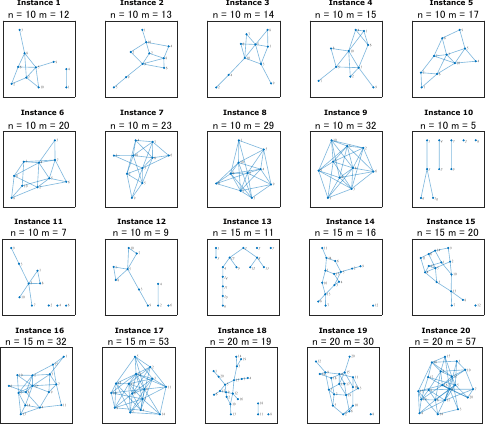}
\caption{Graph instances used in this study.}
\label{maps}
\vspace{-0.5cm}
\end{figure*}

\section{Computational Experiments}


To evaluate the performance of the proposed approach, we conducted computational experiments using diverse graph topologies, configuration of nodes and edges, with varying degrees of sparsity. Fig. \ref{maps} shows the set of graph instances. For ease of reference, plots correspond to arbitrary ordering/labeling of nodes (in the range of 10 to 20), considering the modeling of labeled trees as well. The key motivation for using the settings in Fig. \ref{maps} is due to our interest in finding compact representations for multiagent communication networks, where the number of agents is expected to be distributed in indoor environments (e.g., robots, drones, sensors distributed in the map). As such, exploring the compact representation of graphs is desirable to communicate and share the topology of the graph among members of a robot swarm. The reader may also note that some graphs in Fig. \ref{maps} are not connected to the entire network. This observation (and situation) is especially relevant when agents (nodes) are independent of the group (swarm); thus, communication and information gathering is not essential at all times.

To evaluate the feasibility in attaining quick convergence towards graph representations with smaller integers, we use a representative class of optimization algorithms derived from Differential Evolution, which is a population-based gradient-free optimization algorithm extending the difference of vectors:

\begin{enumerate}
\item \small \textsf{DESPS}: Differential Evolution with Successful Parent Selection\cite{desps}
\item \textsf{OBDE}: Differential Evolution with Opposition-based Learning\cite{rahnamayan2008opposition}
\item \textsf{JADE}: Adaptive Differential Evolution External Archive\cite{zhang2009jade}
\item \textsf{DCMAEA}: Differential Covariance Matrix Adaptation\cite{dcmaea12}
\item \textsf{DERAND}: DE/rand/1/bin Strategy\cite{storn1997differential},
\item \textsf{DEBEST}: DE/best/1/bin Strategy\cite{storn1997differential},
\item \textsf{RBDE}: Rank-based Differential Evolution\cite{rbde},
\item \textsf{DESIM}: Differential Evolution with Similarity Based Mutation\cite{segredo2018novel}.
\end{enumerate}

The key motivation of using the above set is due to our interest in evaluating distinct forms of selection pressure, parameter adaptation, exploration and explotation mechanisms during search. As for dimensionality of the problem, since $x \in \mathbf{F}(n)$, the dimensionality is related to $n$ due to node labelings in the factorial search space. For simplicity and without loss of generality, we used the factoradic representation to encode solutions in the factorial search space. As for algorithm parameters, we used the probability of crossover $CR = 0.5$, scaling factor $F = 0.7$, population sizes $POP = 10$, the bias term in \textsf{RBDE} $\rho = 3$, and the termination criterion is $\Omega = 1000$ function evaluations. In \textsf{DESIM}, the similarity is based on the Euclidean distance, the learning parameters $F$ and $CR$ are adapted using \textsf{JADE}\cite{zhang2009jade}, and initialization uses Opposition Based Learning\cite{rahnamayan2008opposition}.

Also, due to the stochastic nature of the above mentioned metaheuristics, 10 independent runs were evaluated for each algorithm and each configuration. Other parameters followed the suggested values of the above-mentioned references. The fine-tuning of optimization parameters is out of the scope of this paper. The key motivations for using the above parameters are as follows: Crossover probability with $CR = 0.5$ implements the equal importance and consideration to historical search directions up to the current number of iterations $t$. Small population size $NP = 10$ and number of evaluations up to 1000 allow evaluating the (frontier) performance of the gradient-free algorithms under tight computational budgets.

Computational experiments were conducted in Matlab 2018b. Since computing the number $g$ that represents a graph involves the summation of binomial coefficients, we rather use the logarithm of $g$ to efficiently compute the reduction of large binomial coefficients. In this paper, the computation of logarithmic function of the number $g$ is realized through the parallel reductions in a GPU (NVIDIA GeForce GTX TITAN X). The reason of using a GPU is due to the efficiency in the reduction scheme; however, since we are constrained by the GPU hardware, we can only explore the possibility of computing the logarithms within the bound of \verb"realmax" in Matlab ($1.8 \times 10^{308}$). As such, exploring the compact number-based representations for large graphs is out of the scope of this paper (due to hardware limitations) and is left for future work in our agenda (potentially through a parallel distributed computing environment).

\subsection{Results and Discussion}


Fig. \ref{conv} shows the mean convergence performance through distinct graph instances and independent runs. Also, the x-axis of each plot in Fig. \ref{conv} shows the number of function evaluations, whereas the y-axis shows the value of the objective function $L$. By observing Fig. \ref{conv}, we note the following facts: (1) convergence and stagnation in flat regions often occur after 200 function evaluations with graphs with ten nodes and 20 edges, whereas convergence occurs after 400 functions in other graph instances; (2) the minimal objective values corresponding to each graph instance differ from one to another. It is possible to observe the pattern that a large (small) number of edges implies higher (lower) objective values. This is due to the fact that graphs encode dependencies; as such, more information will be necessary to encode graphs with a larger number of edges, and the integer number representing the graph will be larger due to Eq. \ref{comnum} depending on $m$ on the summation component.

\begin{figure}[t!]
\centering
\begin{tblr}{
    colspec = {Q[l]},
    rowsep = 3pt,
    colsep = 1pt,
    }
    \stackon{\includegraphics[width=0.24\columnwidth]{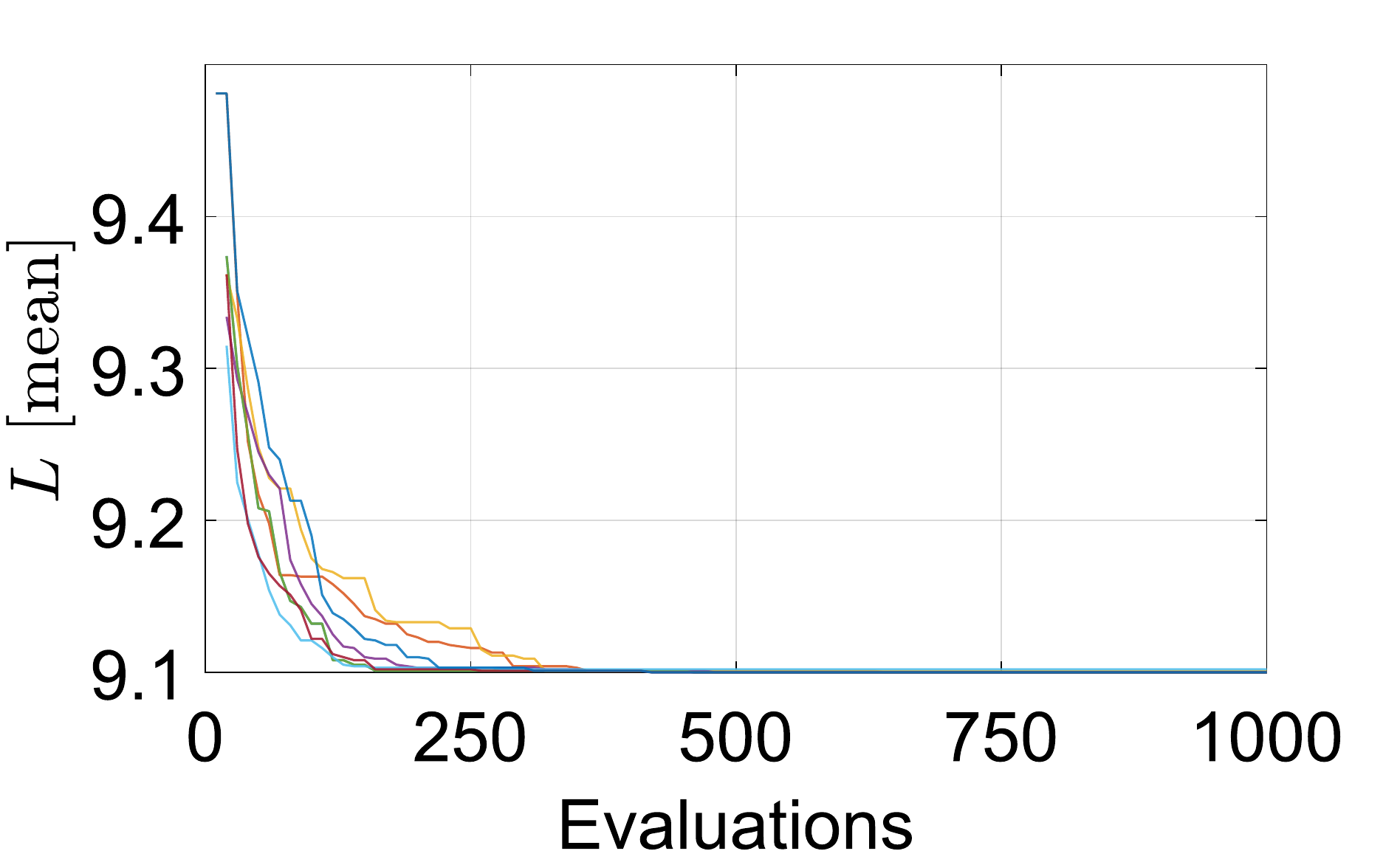}}{\scriptsize Instance 1}
    \stackon{\includegraphics[width=0.24\columnwidth]{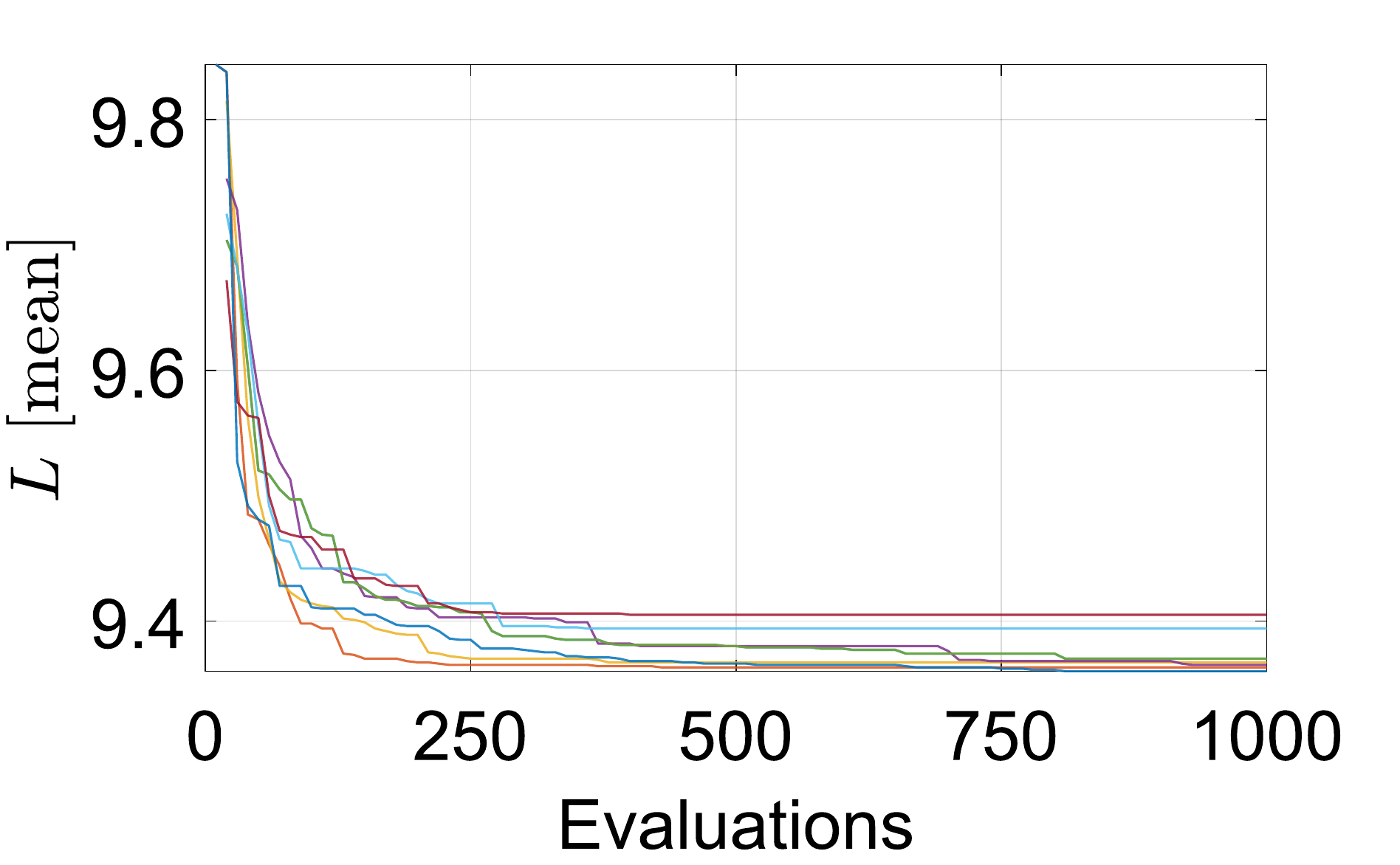}}{\scriptsize Instance 2}
    \stackon{\includegraphics[width=0.24\columnwidth]{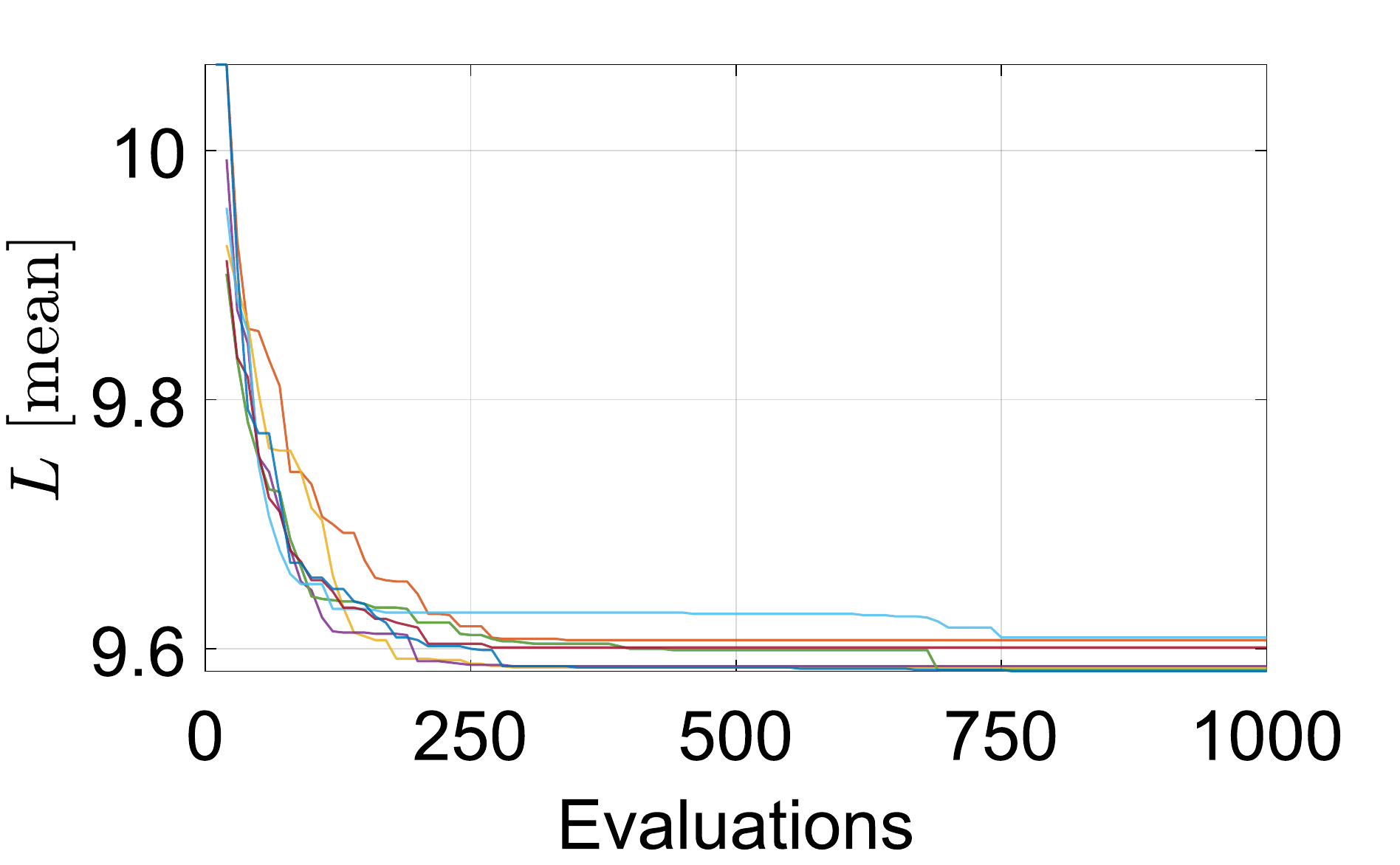}}{\scriptsize Instance 3}
    \stackon{\includegraphics[width=0.24\columnwidth]{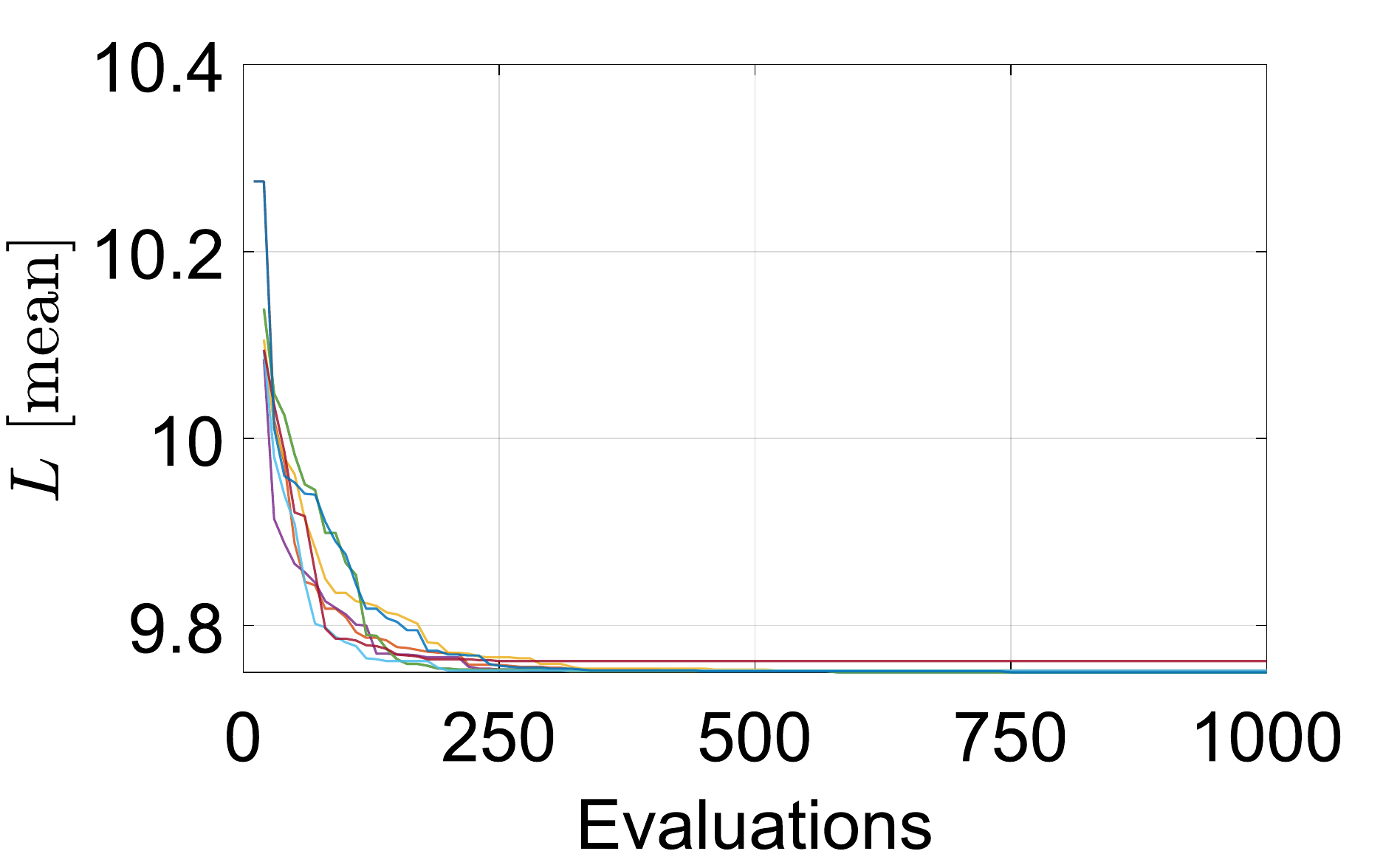}}{\scriptsize Instance 4}
	\\
    \stackon{\includegraphics[width=0.24\columnwidth]{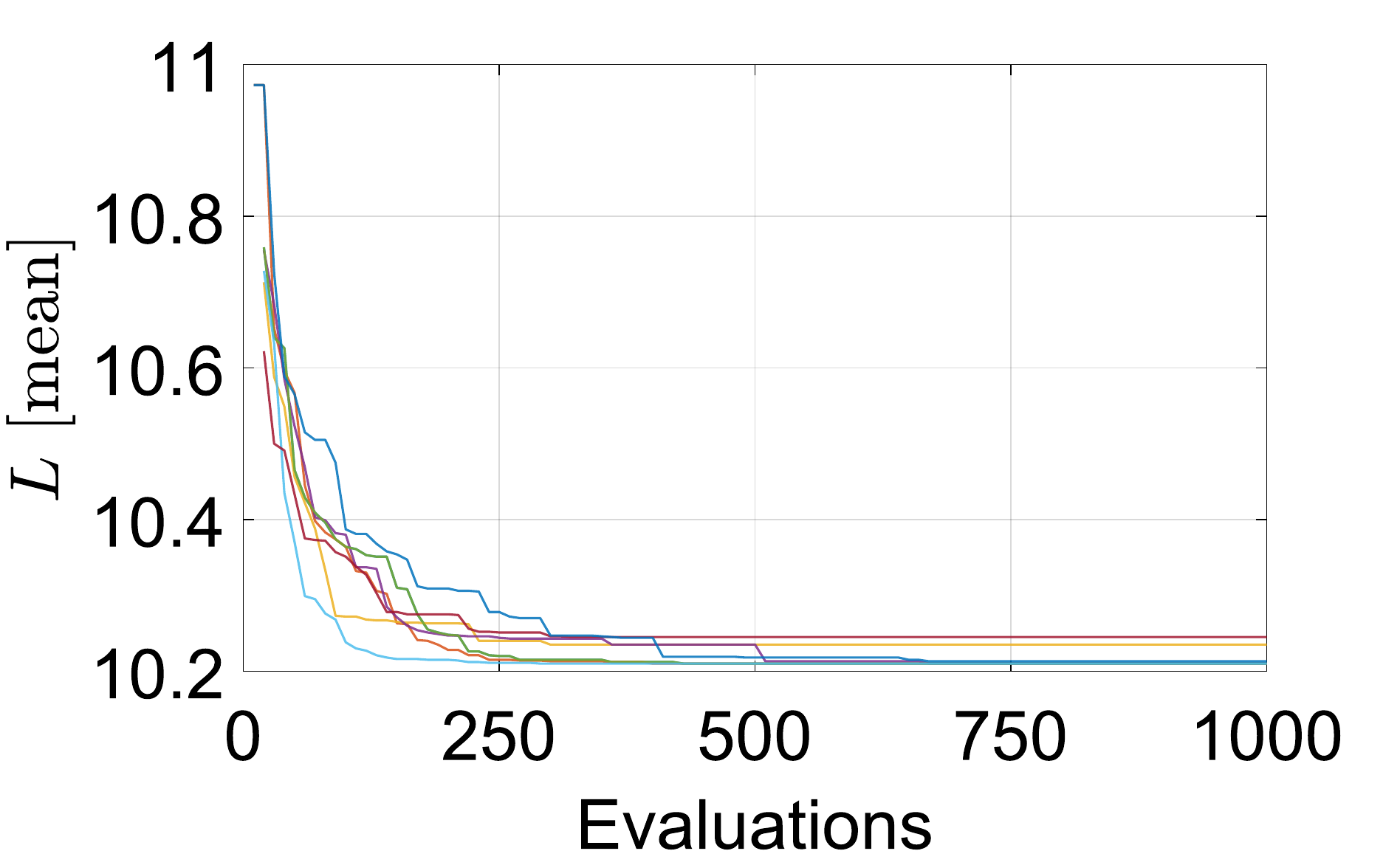}}{\scriptsize Instance 5}
    \stackon{\includegraphics[width=0.24\columnwidth]{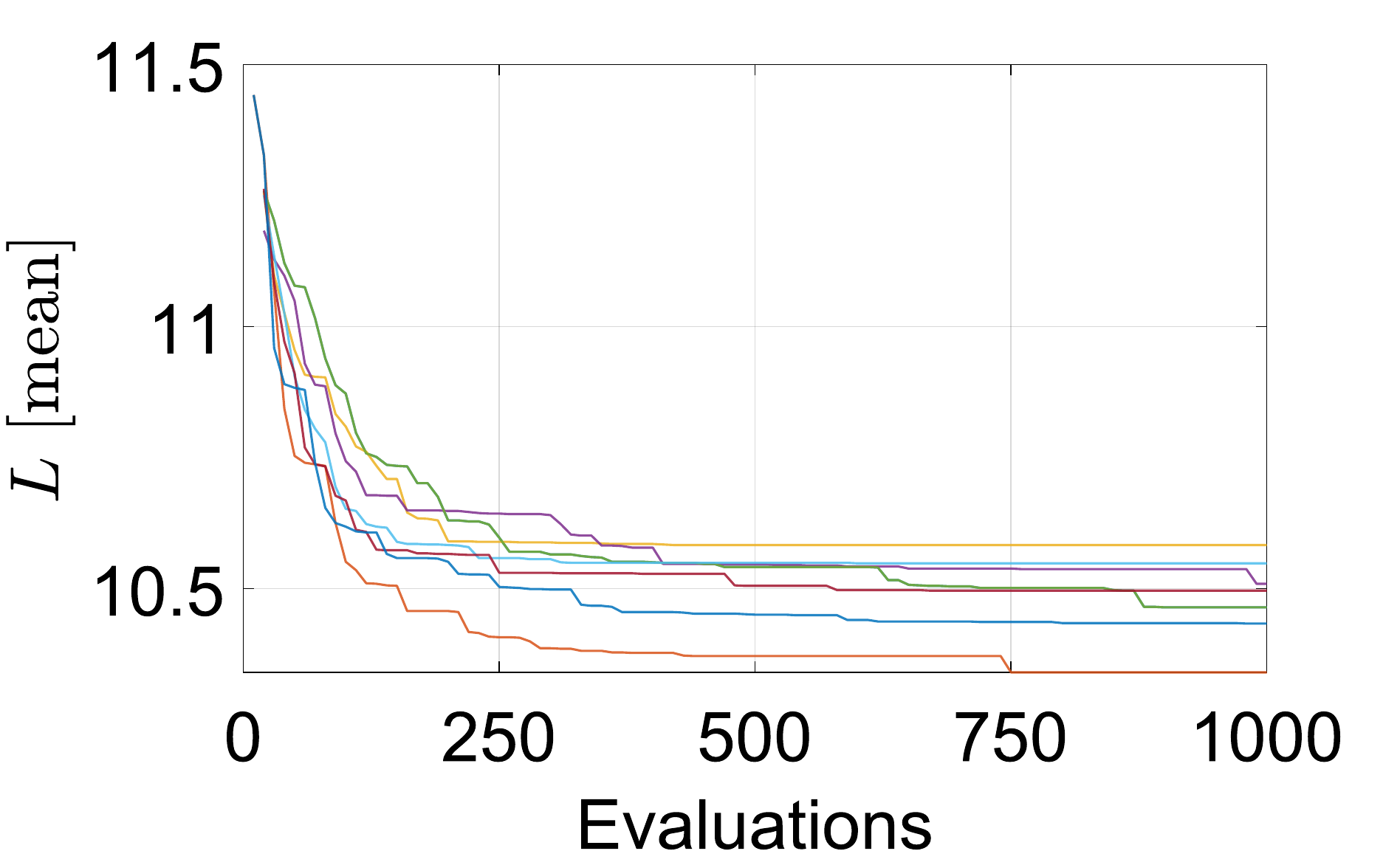}}{\scriptsize Instance 6}
    \stackon{\includegraphics[width=0.24\columnwidth]{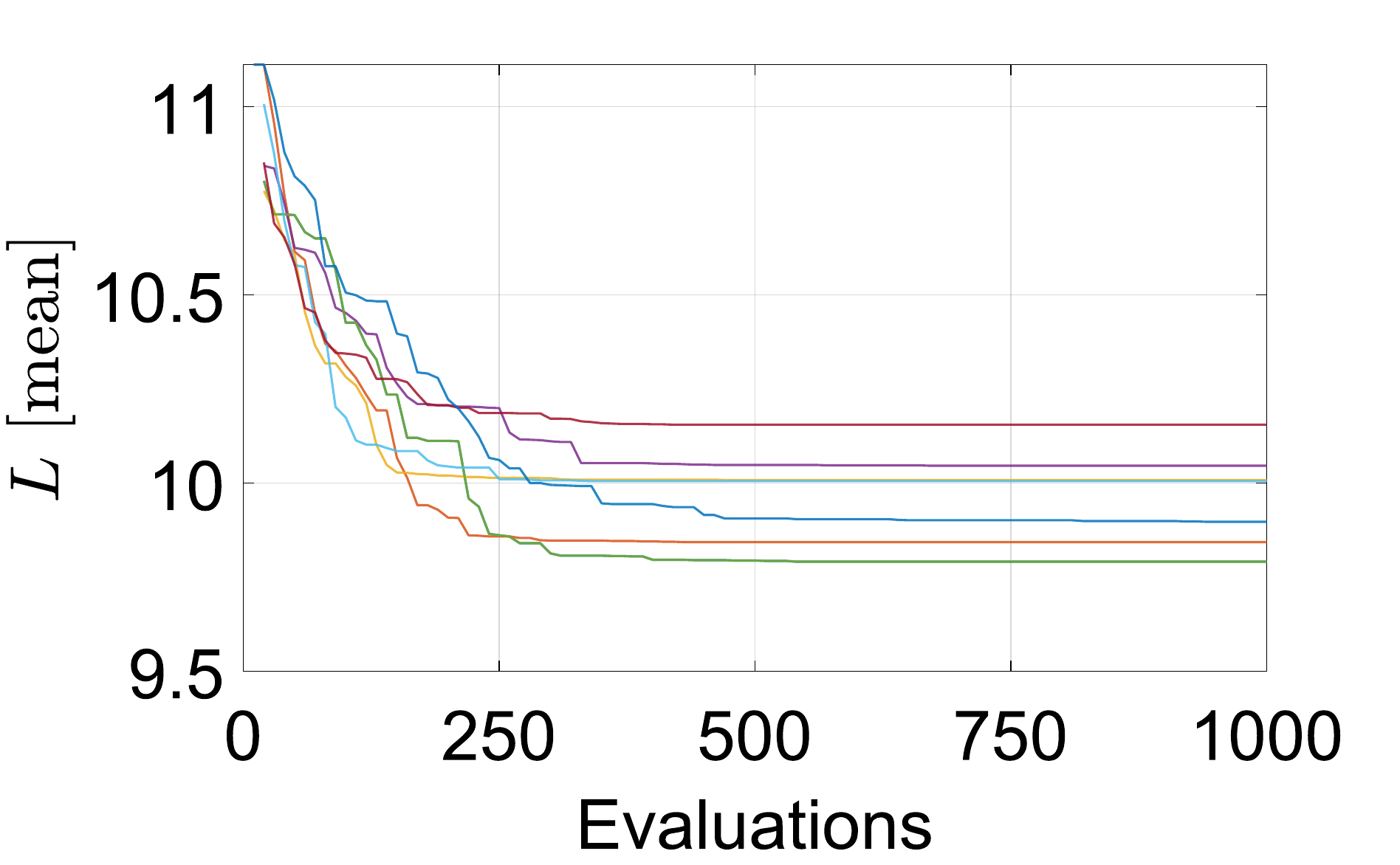}}{\scriptsize Instance 7}
    \stackon{\includegraphics[width=0.24\columnwidth]{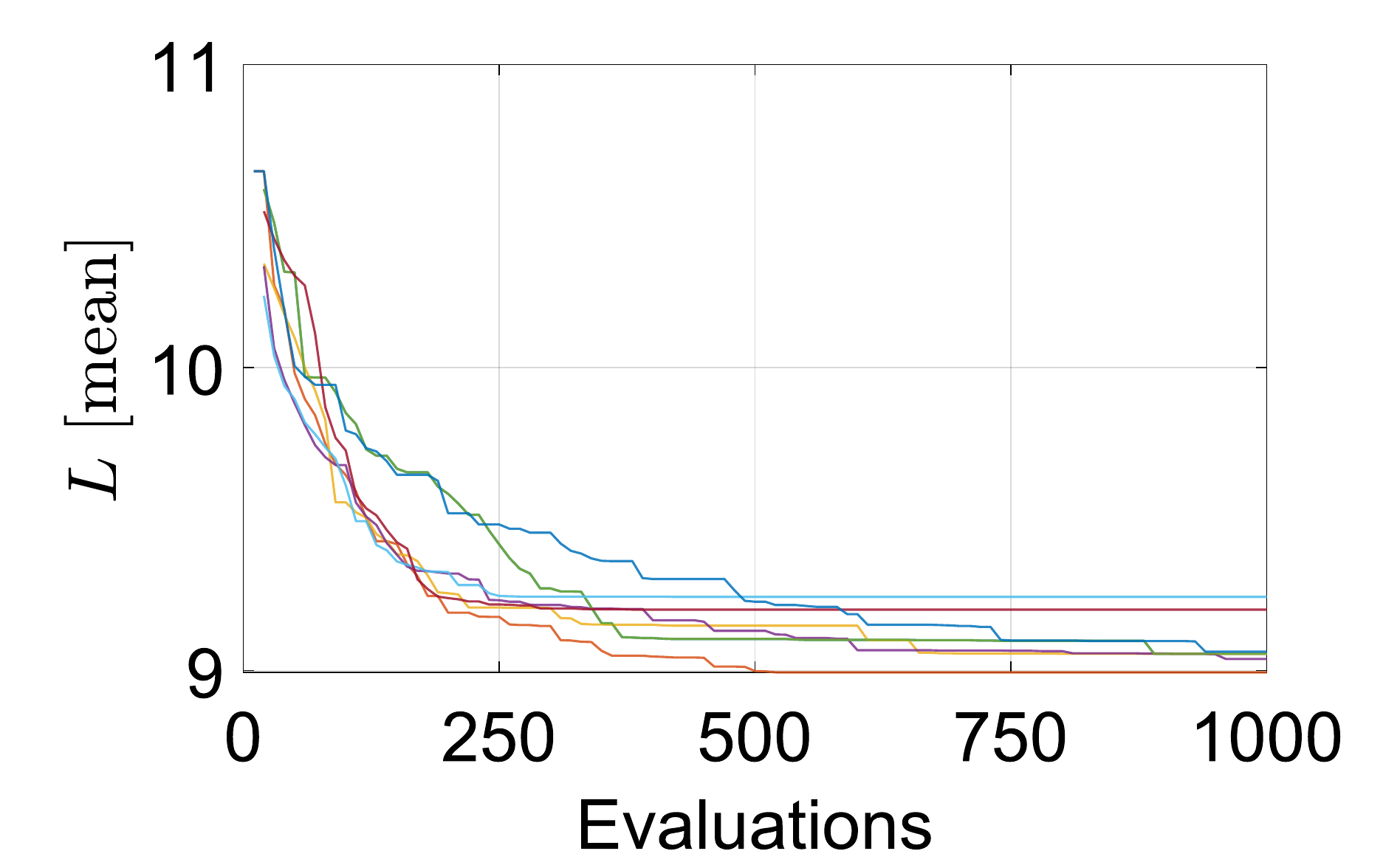}}{\scriptsize Instance 8}
    \\
    \stackon{\includegraphics[width=0.24\columnwidth]{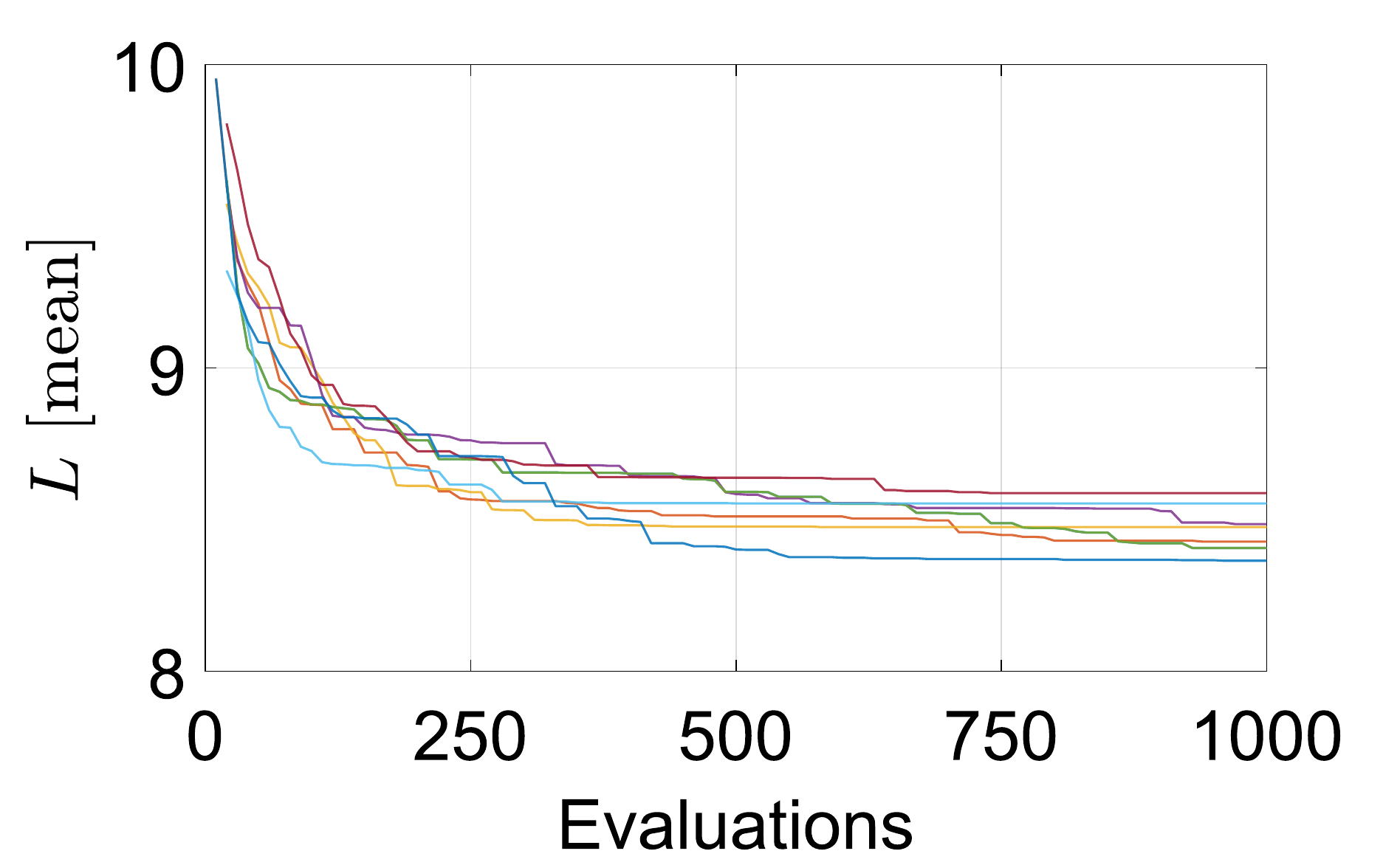}}{\scriptsize Instance 9}
    \stackon{\includegraphics[width=0.24\columnwidth]{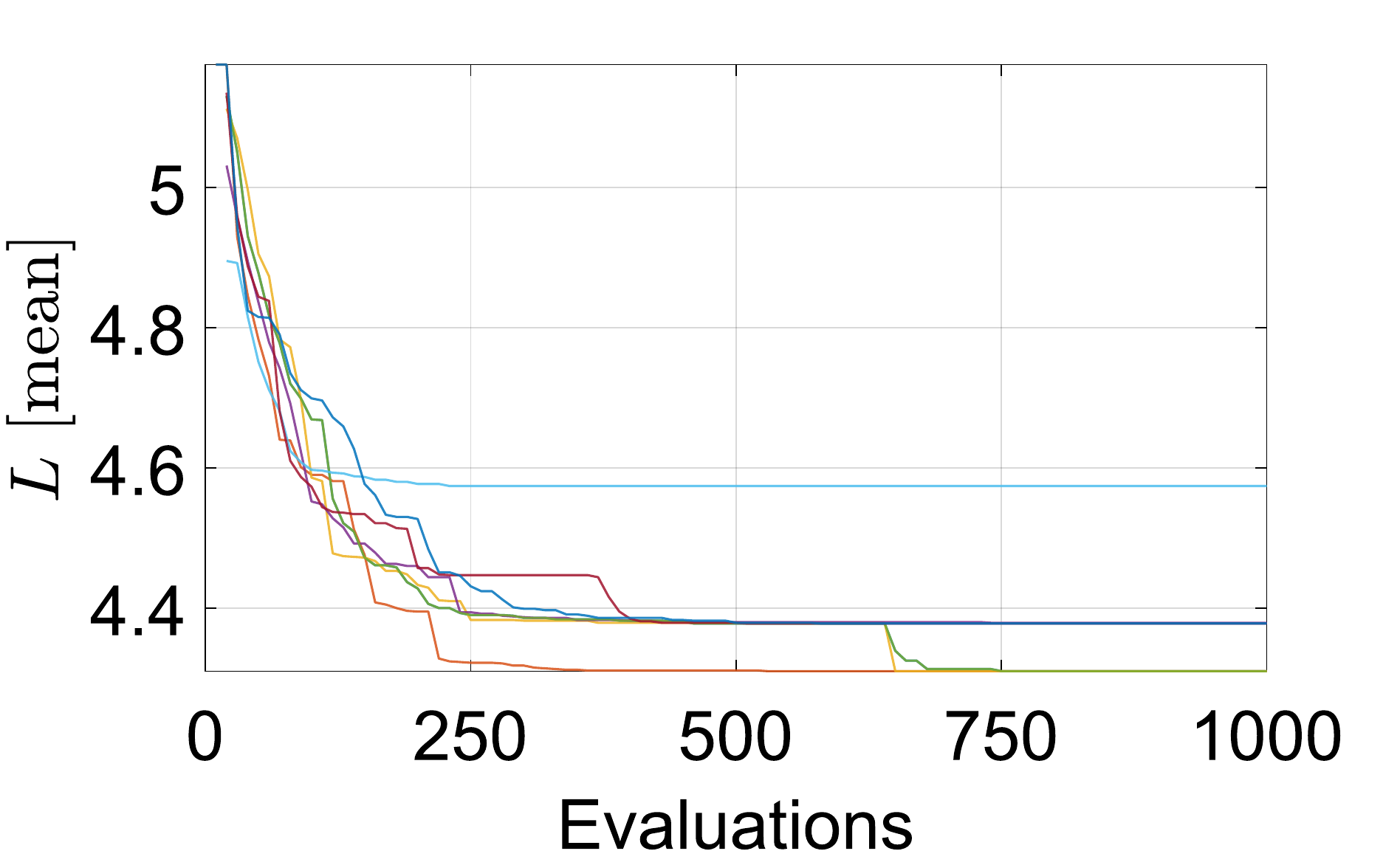}}{\scriptsize Instance 10}
    \stackon{\includegraphics[width=0.24\columnwidth]{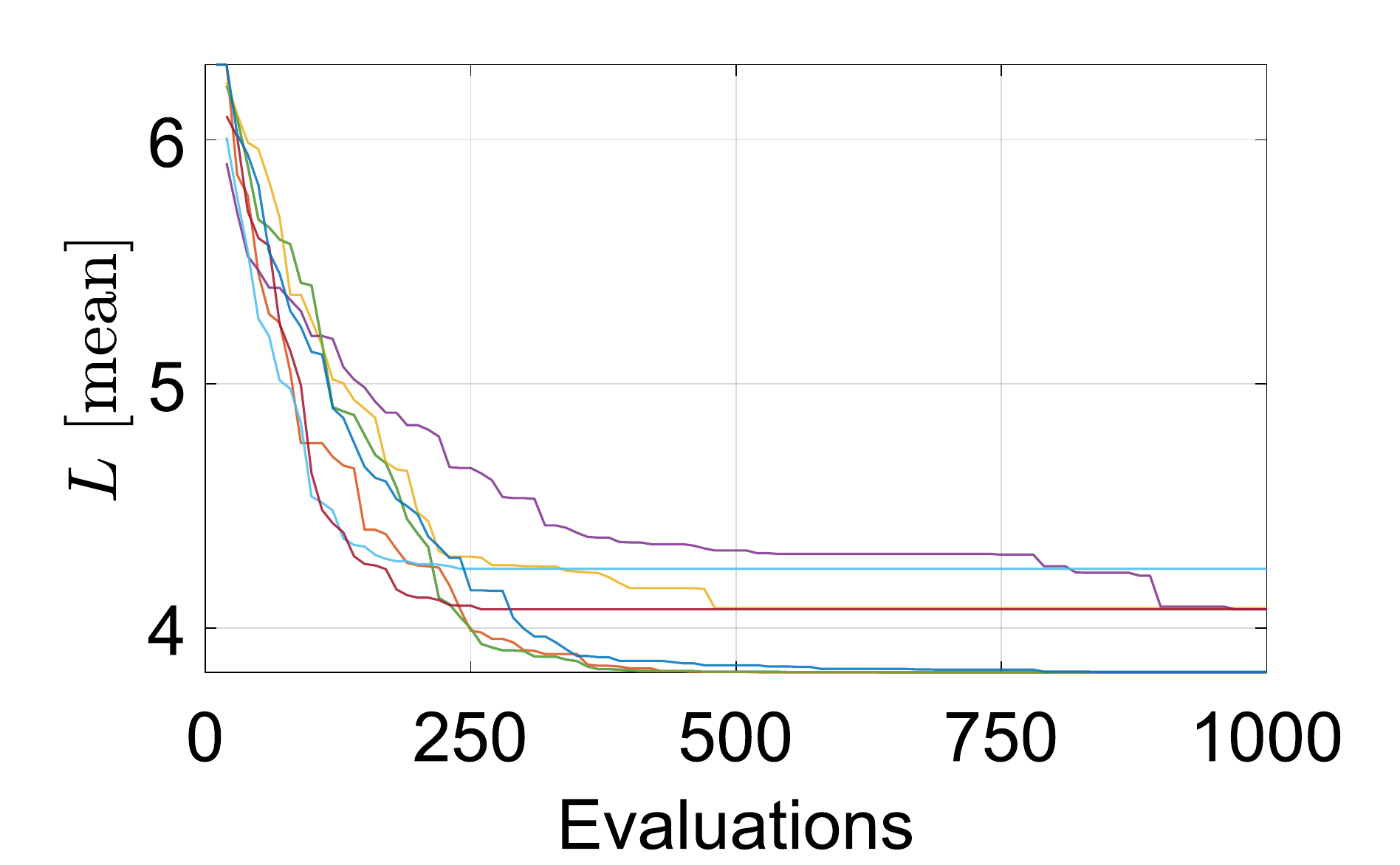}}{\scriptsize Instance 11}
    \stackon{\includegraphics[width=0.24\columnwidth]{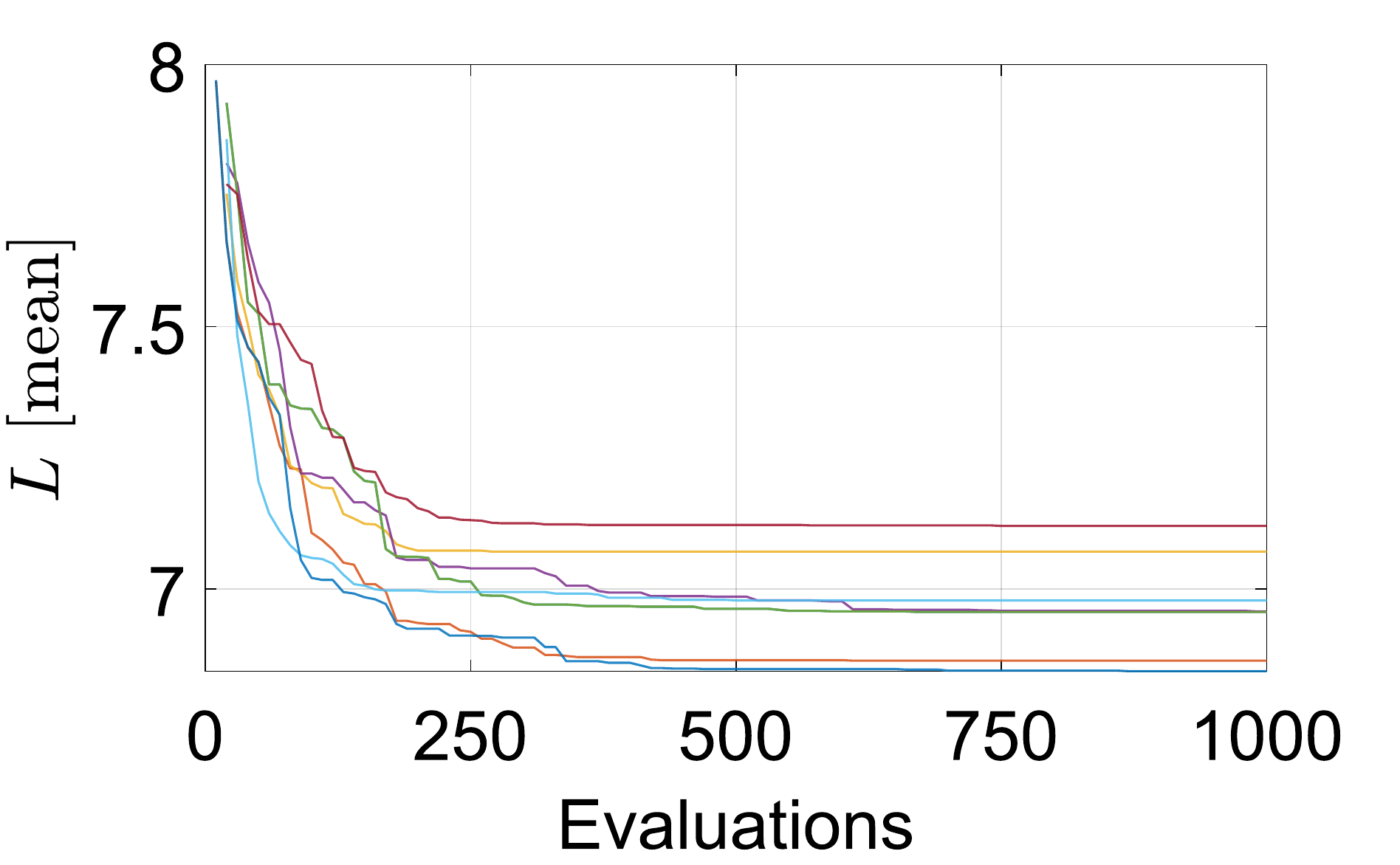}}{\scriptsize Instance 12}
    \\
    \stackon{\includegraphics[width=0.24\columnwidth]{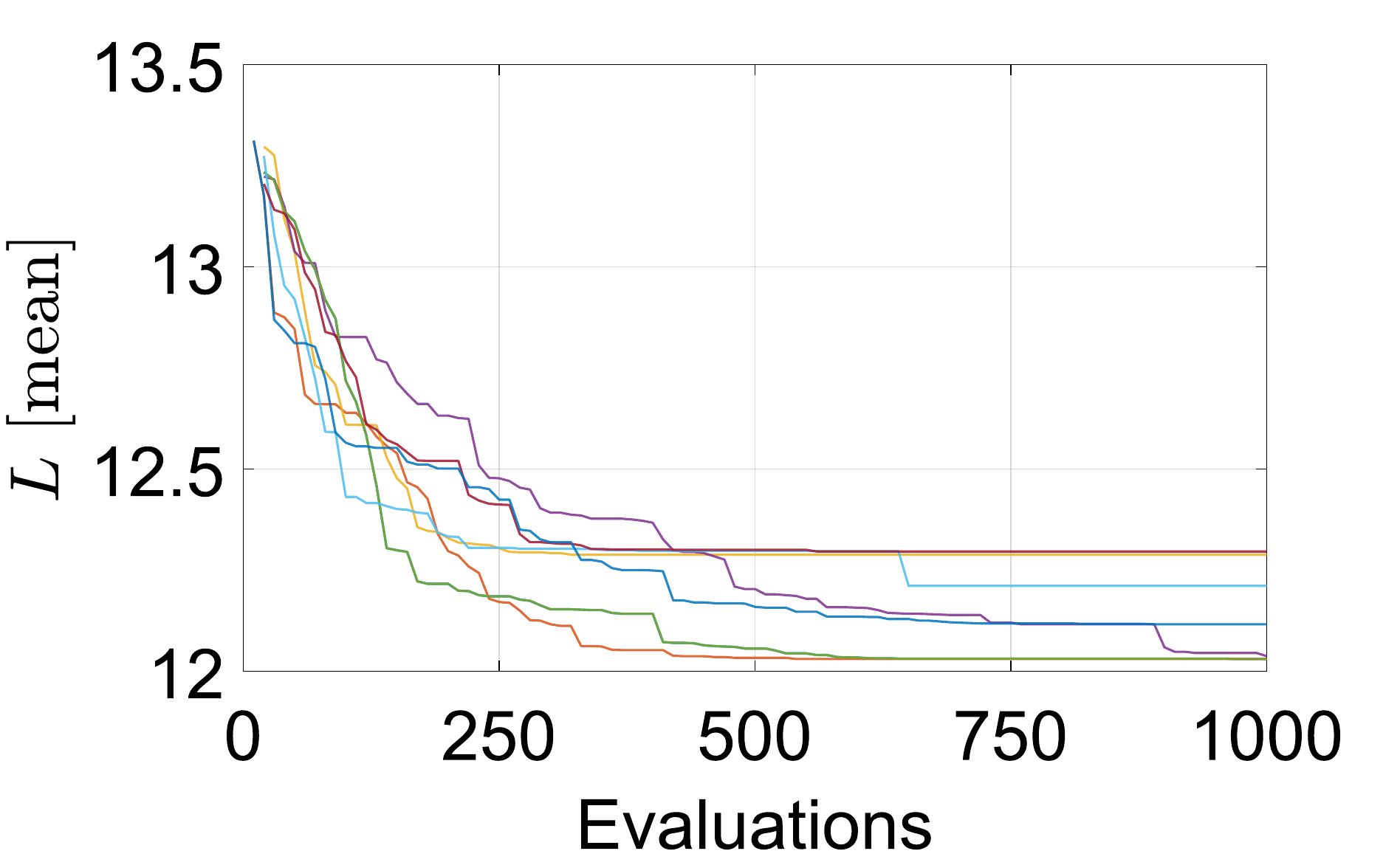}}{\scriptsize Instance 13}
    \stackon{\includegraphics[width=0.24\columnwidth]{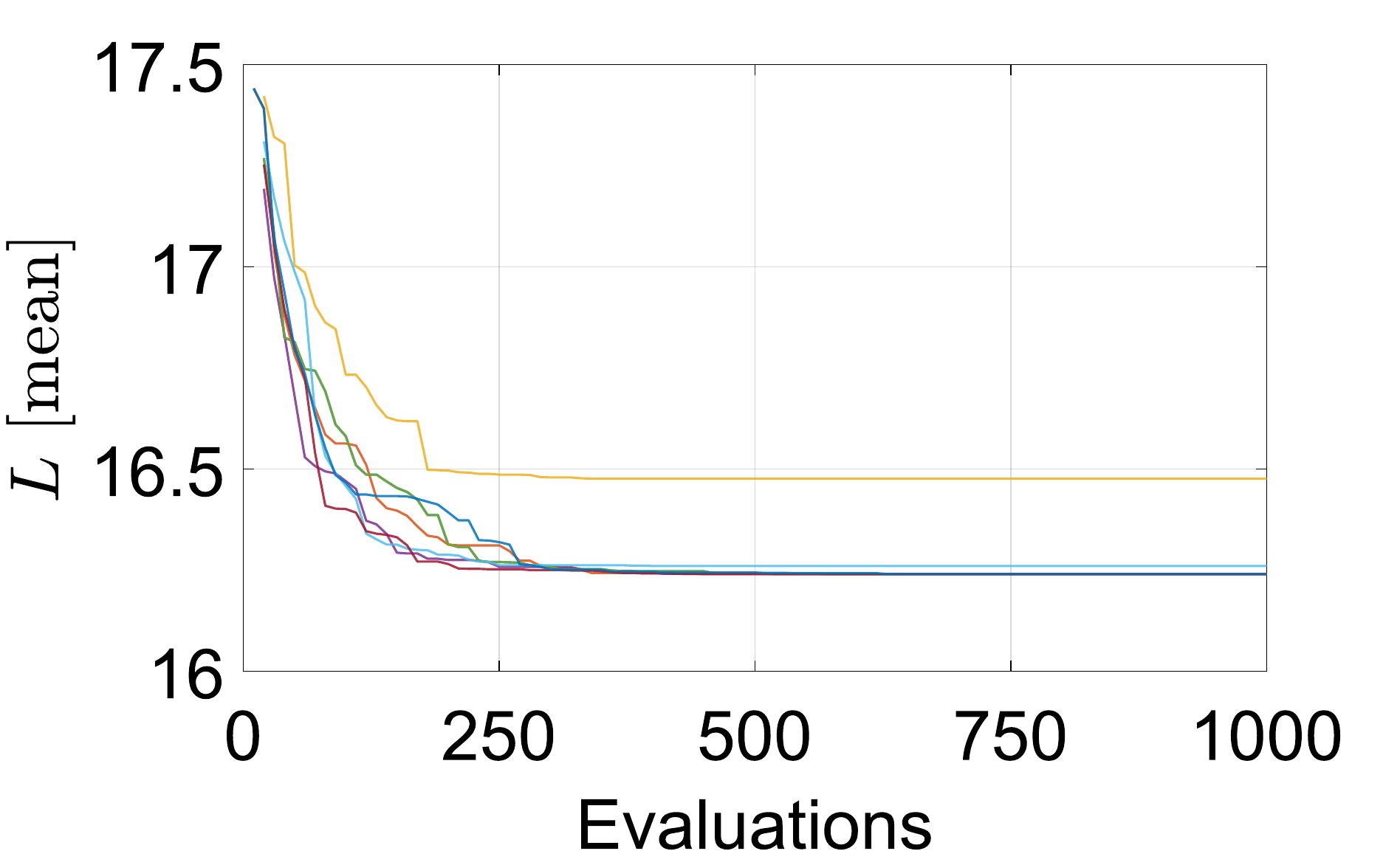}}{\scriptsize Instance 14}
    \stackon{\includegraphics[width=0.24\columnwidth]{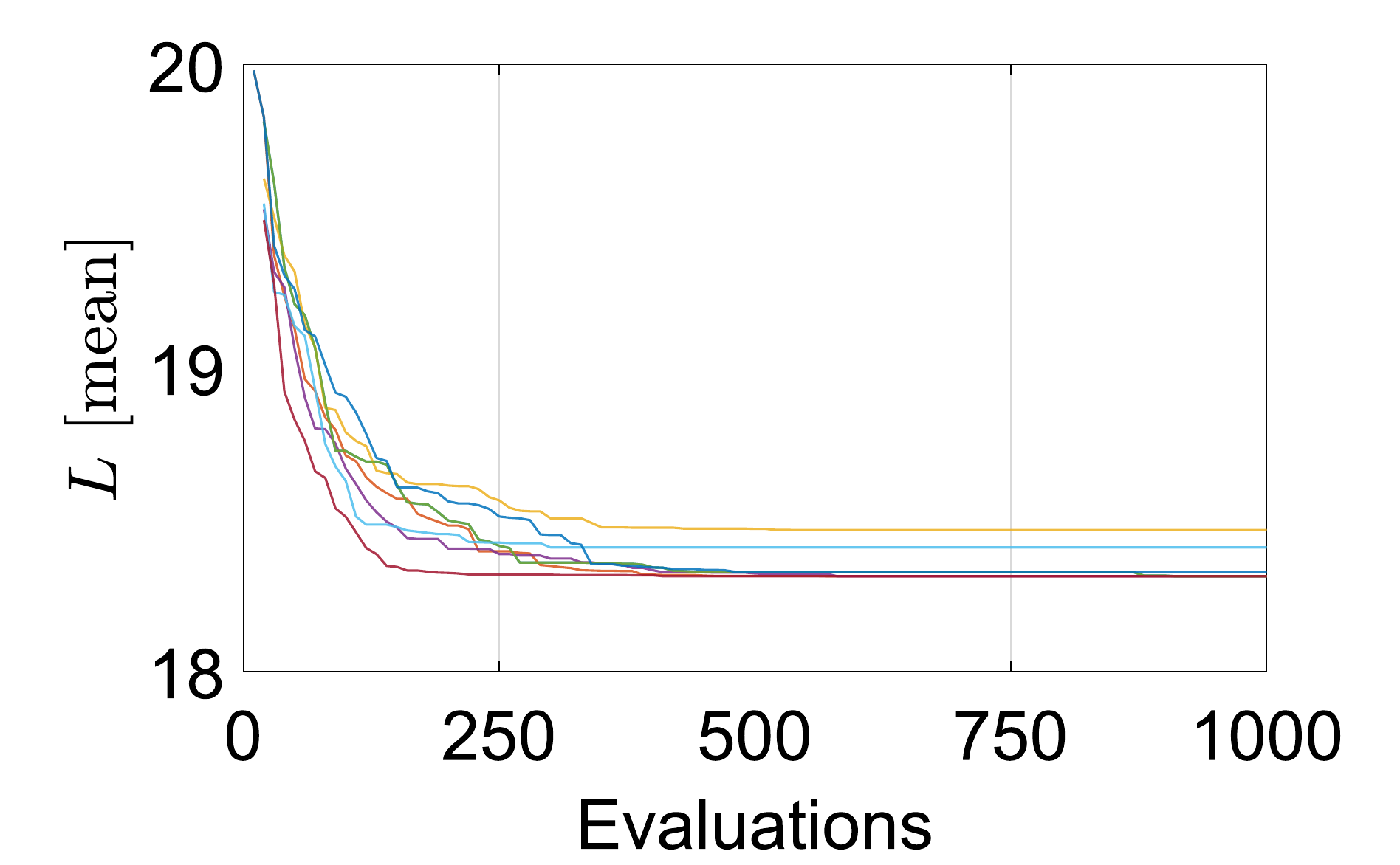}}{\scriptsize Instance 15}
    \stackon{\includegraphics[width=0.24\columnwidth]{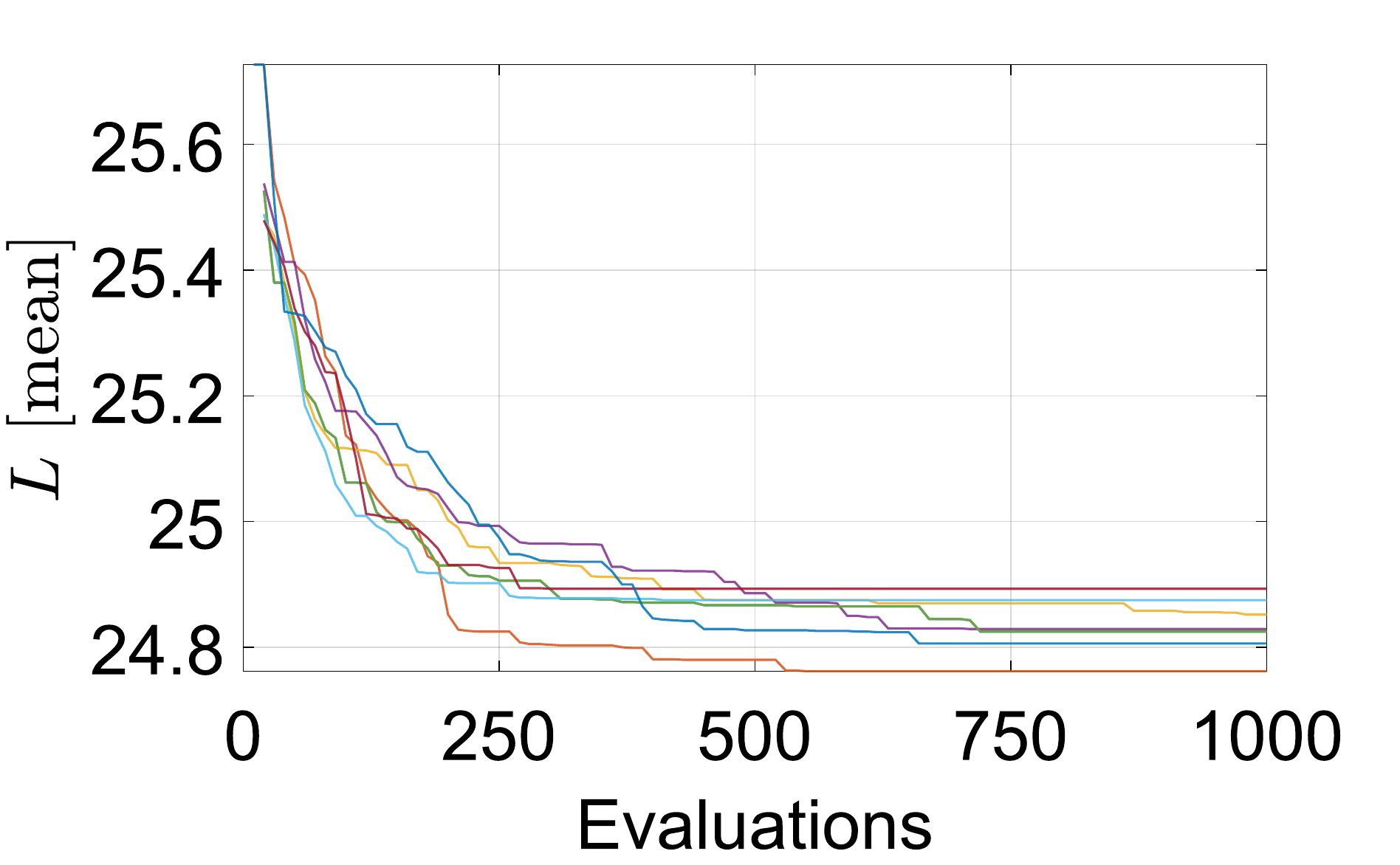}}{\scriptsize Instance 16}
    \\
    \stackon{\includegraphics[width=0.24\columnwidth]{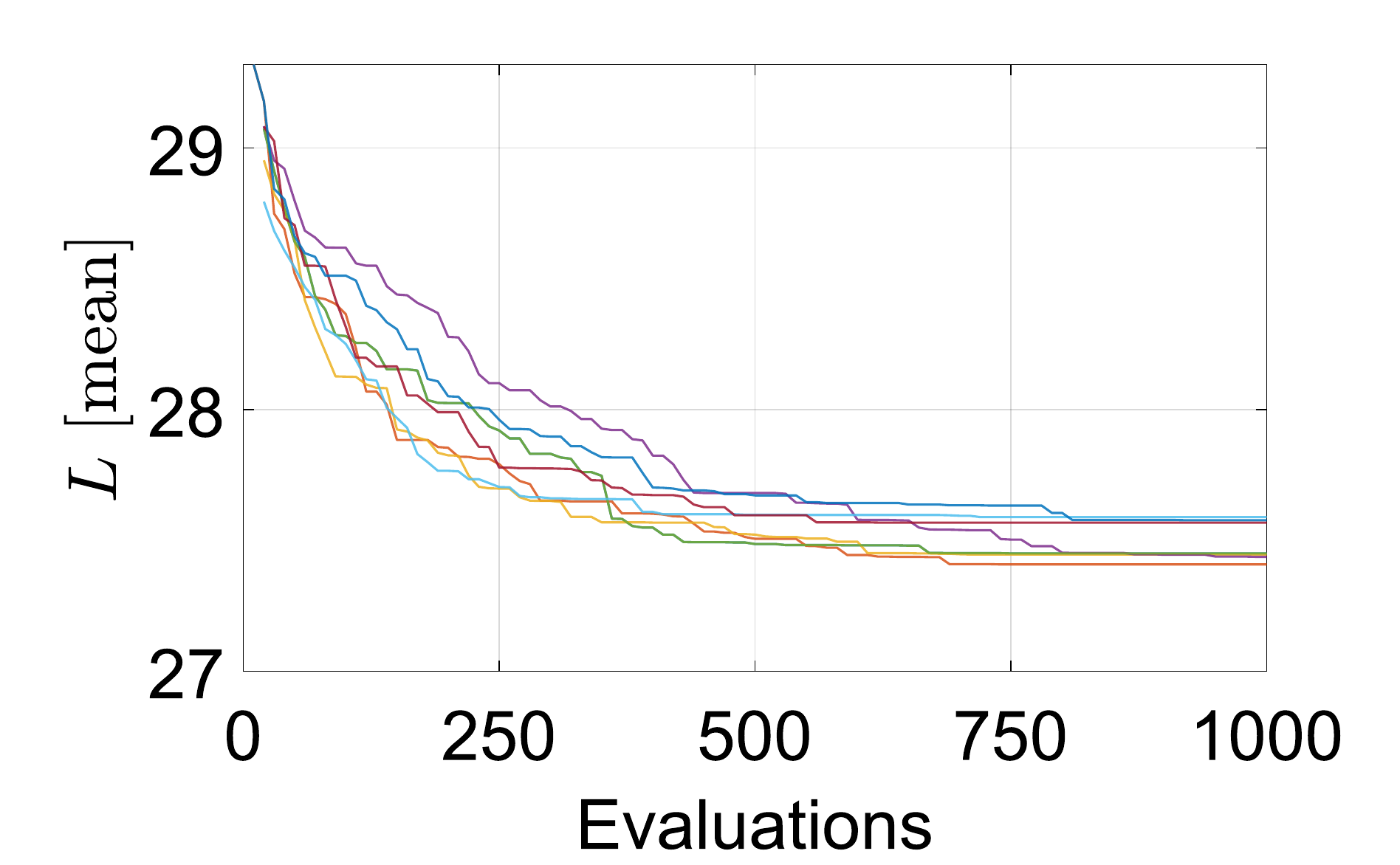}}{\scriptsize Instance 17}
    \stackon{\includegraphics[width=0.24\columnwidth]{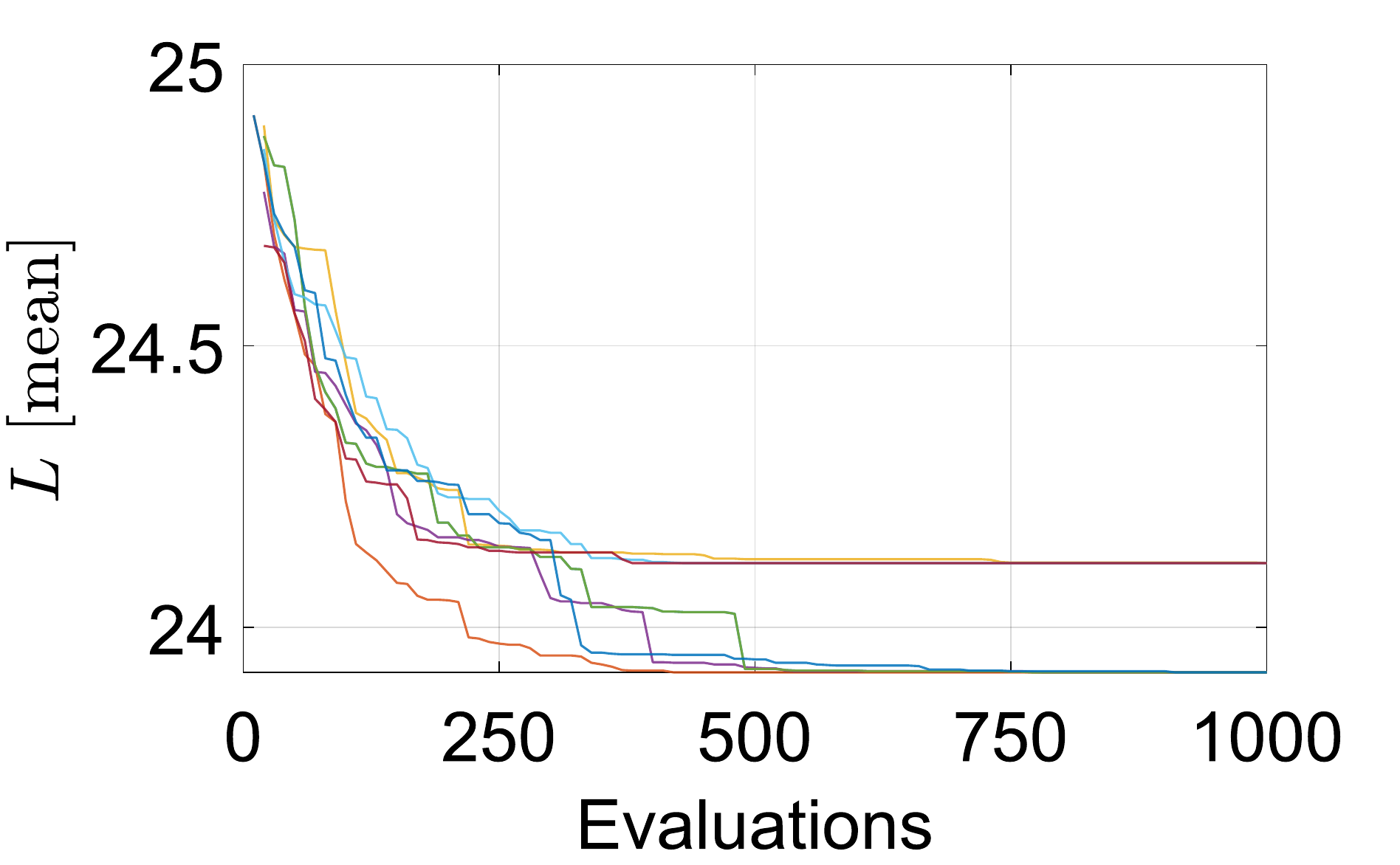}}{\scriptsize Instance 18}
    \stackon{\includegraphics[width=0.24\columnwidth]{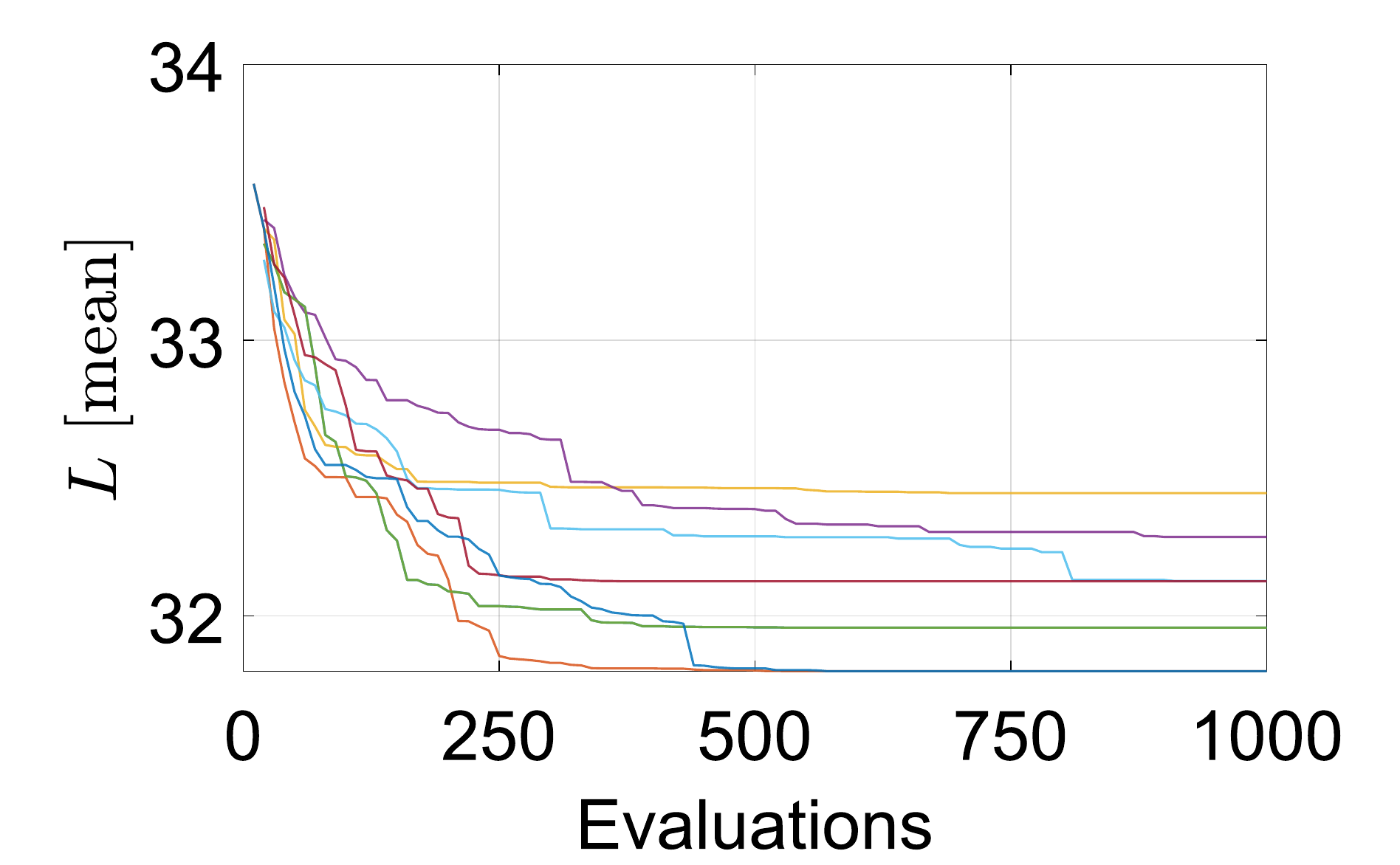}}{\scriptsize Instance 19}
    \stackon{\includegraphics[width=0.24\columnwidth]{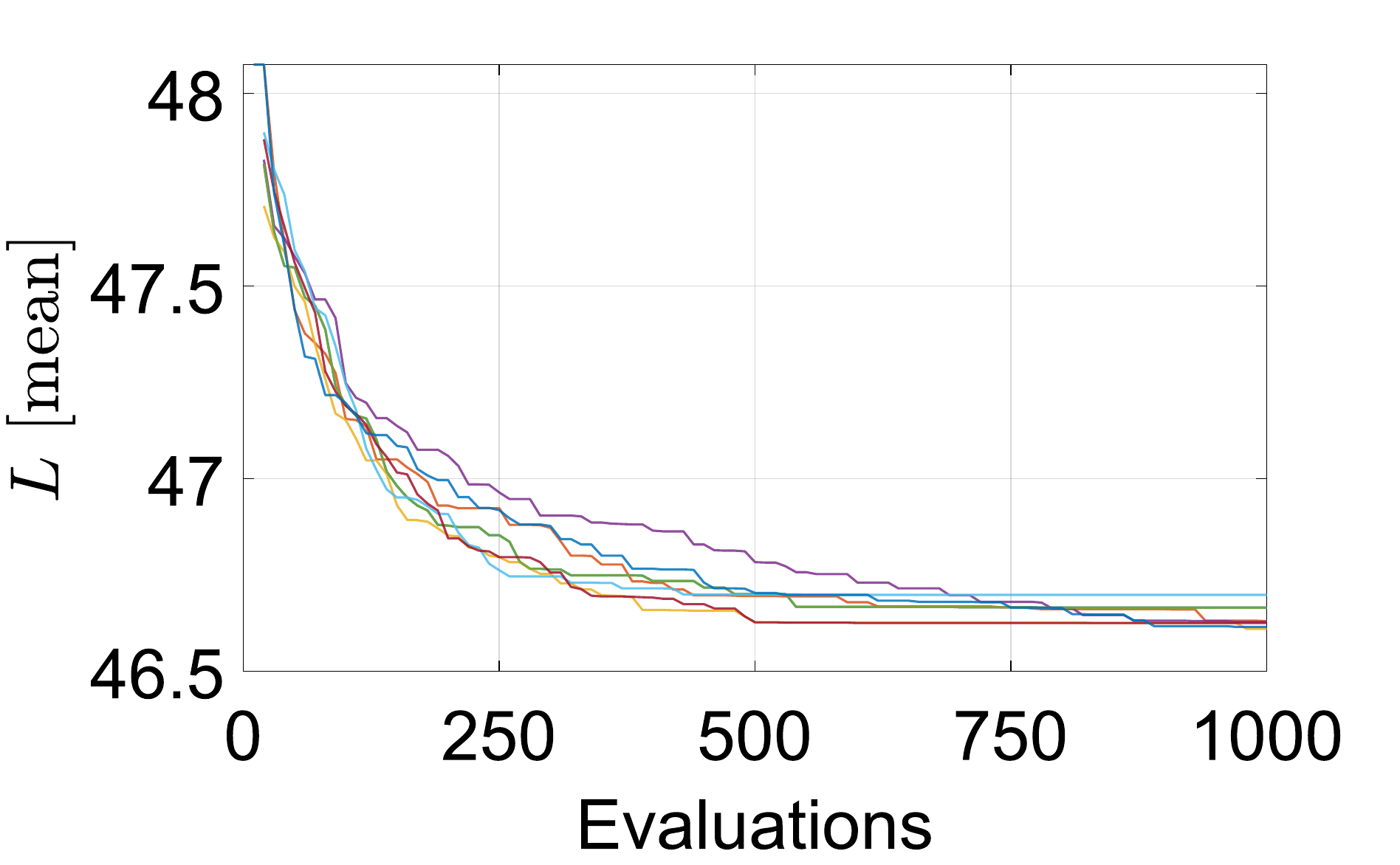}}{\scriptsize Instance 20}
    \\
    \centering
    \includegraphics[width=0.98\columnwidth]{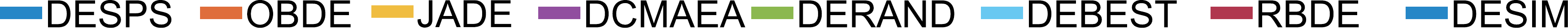}       
\end{tblr}
\caption{Mean convergence of the objective function $L$ across independent runs. }
\label{conv}
\vspace{-0.5cm}
\end{figure}

Among all studied graph instances, the objective value of graph instance 10-11 is lower compared to the objective values of other graph instances. Yet, for all cases, it is possible to attain lower objective values for all cases. Overall cases, the algorithms relying on exploration, such as DERAND, and the algorithm relying on heavy exploration during the early stages of the search, such as DESIM show better convergence performance. This observation is due to the noisy landscape of the problem, such as those depicted in Fig. 3 and Fig. 4. Despite the large factorial search space, the convergence figures show that it is possible to find potential areas of low objective values using few function evaluations.

\begin{figure*}[t]
\centering
\includegraphics[width=0.98\textwidth]{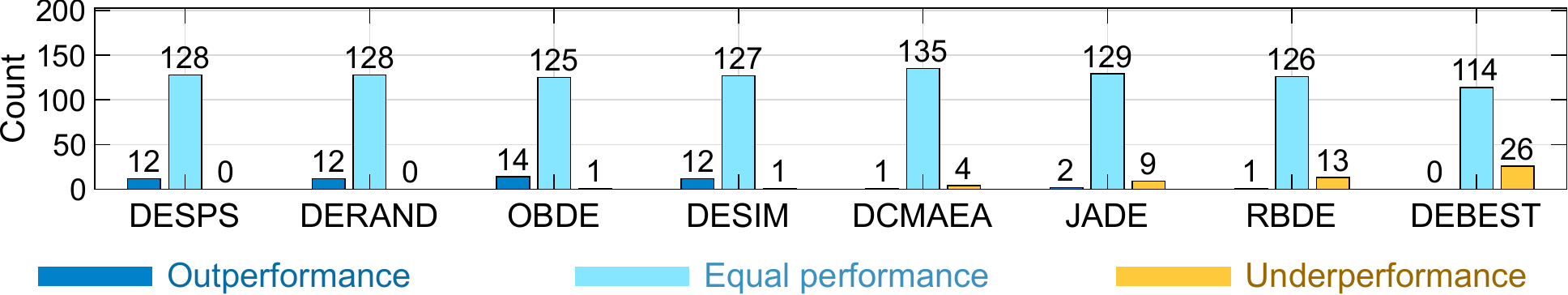}
\caption{Number of cases of outperformance, equal performance and underperformance derived from pair-wise Wilcoxon tests at 5\% significance level.}
\label{count}
\vspace{-0.5cm}
\end{figure*}


The observations above show it is possible to find convergence regions quickly and that exploration strategies are relevant for better performance. To show the comparison of the performance of the algorithms across graph instances, we performed statistical test at 5\% significance level (Wilcoxon). Here, our goal is to compute instances in which an algorithm is significantly better(+), equal(=), or worse(-) compared to another algorithm. As such, Fig. \ref{count} shows the overall performance comparison among algorithms based on the statistical significance. Here, the bars in the plot show the number of times an algorithm is significantly better (outperformance), equal (equal performance), or worse (underperformance) compared to other algorithms across all graph instances overall independent runs. For clarity of exposition, the counts (numbers depicted over each bar) are computed by accumulating the number of events in which [+], [=], and [-] occur from the Wilcoxon test. By observing Fig. \ref{count}, we note that algorithms with an explicit exploration strategy outperform other algorithms. For instance, DESPS, DERAND, OBDE and DESIM outperform other algorithms in 12-14 problem instances. Also, algorithms with exploitation strategies such as DEBEST and RBDE perform equally in most problem scenarios, yet underperform in 13-26 instances. 


The above results show that using gradient-free heuristics can enable the search for compact integer representations of graphs using few evaluations, which may find use in modeling and sampling graphs and networks with utmost efficiency. The future work will aim at studying the tailored optimization algorithms extending adaptation and explorative strategies, and the modular schemes for large graphs\cite{compsac18mod}. It also remains to be answered whether different labeling strategies than the revolving door\cite{smc14} are able to achieve smaller values of the objective function (smaller numbers representing graphs). Furthermore, it remains to be answered how to achieve minimal number representation for weighted graphs, directed graphs\cite{bigcomp17}, loopy graphs\cite{ictai17}, and other relevant graph architectures. On the application side, we aim at using the insights from this study to extrapolate towards large-scale minimal networks in the plane\cite{evocop21,ictai18star}, and the further works in network design and optimization.


\section{Conclusion}

We have studied the feasibility of gradient-free population-based optimization heuristics based on Differential Evolution variants for minimal integer representations of graphs. Our computational experiments using graph instances with varying degrees of sparsity and eight optimization algorithms with distinct forms of adaptation and exploration-exploitation have shown the merit of exploration-oriented strategies to attain better convergence with fewer function evaluations. Investigating the role of adaptation, exploration and modular strategies in large graphs, and evaluating other forms of graph encoding strategies is left for future work. Our results have the potential to elucidate new number-based approaches for graph representation, network design and optimization.

\section*{Acknowledgment}

This work was supported by JSPS KAKENHI Grant Number 20K11998.


\bibliographystyle{unsrtnat}
\bibliography{mybiblio}

\end{document}